\documentclass[a4paper,11pt]{article}
\usepackage{graphicx,subcaption}
\usepackage{subcaption}
\usepackage{caption}
\usepackage{pos}

\DeclareMathOperator{\ReN}{Re}
\DeclareMathOperator{\ImN}{Im}

\DeclareMathOperator{\tr}{Tr}

\newcommand{\dif}[1]{\mathrm{d} #1 }

\begin{document}
	\title{Mean field approximation for effective theories of lattice QCD}
	\author*{Christoph Konrad}
	\author{Owe Philipsen}
	\author{Jonas Scheunert}
	\affiliation{Institute for Theoretical Physics, Goethe-University Frankfurt am Main,\\Max-von-Laue-Str. 1,  60438 Frankfurt am Main, Germany}
	\emailAdd{konrad@itp.uni-frankfurt.de}
	\emailAdd{philipsen@itp.uni-frankfurt.de}
	\abstract{For the exploration of the phase diagram of QCD, effective Polyakov loop theories derived from lattice QCD provide a valuable tool in the heavy quark mass regime. In practice, the evaluation of these theories is complicated by the appearance of long-range and multipoint interaction terms. On the other hand, it is well known that for theories with such kind of interactions mean field approximations can be expected to yield reliable results. Here, we apply this framework to the critical endpoint of the deconfinement transition and results are compared to the literature. This treatment can also be used to investigate the phase diagram at non-zero baryon and isospin chemical potential.}
	\FullConference{The 39th International Symposium on Lattice Field Theory (Lattice 2022), 8th - 13th August 2022, Hoersaalzentrum Poppelsdorf, University of Bonn, Bonn, Germany
	}
	\maketitle
\section{Introduction}
Despite numerous efforts the determination of the phase diagram of QCD is still an ongoing research task. This is due to the sign problem associated to the evaluation of lattice QCD (LQCD) at non-zero chemical potential. To tackle this problem, dimensionally reduced effective theories of LQCD have been derived using the standard Wilson lattice action at arbitrary chemical potential~\cite{Langelage2011, Fromm2012}. The derivation involves a combined expansion in the lattice gauge coupling $\beta$ and the hopping parameter $\kappa$. The degrees of freedom of such theories are Polyakov loops, which allows for cheaper simulations than the mother theory \cite{Neuman2015}, as well as analytical treatments \cite{Glesaaen2016, Kim:2019ykj}. On the downside, the evaluation of the effective theories is complicated by long-range and multipoint interaction terms appearing beyond leading order in the expansion parameters. 

Moving away from the heavy quark limit, the long-range and multipoint interaction terms start to give important contributions to the effective actions. A well-known technique to deal with such scenarios is the mean field approximation, which can be expected to give accurate results, if the interactions become long-range. Further, mean field approximations have already been applied successfully to effective Polyakov loop models in earlier works, also for the case of non-vanishing chemical potential (see e.g. \cite{Fukushima:2006uv, Greensite:2012xv, Rindlisbacher:2015pea}). Therefore, a promising way of extending these works is the treatment of the effective theories within this framework. 

Applied to the effective actions considered in this work, the linearization in the fluctuations around the mean fields (or, equivalently, around the saddle points) implies a loss of the non-linear dependence of the actions on the Polyakov  loops $L_{\mathbf{x}}$ and $L_{\mathbf{x}}^*$. Here, we will find that a more appropriate treatment of local fluctuations will solve this qualitative discrepancy and, consequently, improve the accuracy of the mean field approximation. 
\section{Overview over the effective theories}
Starting with the Wilson gauge action $S_G$ with gauge coupling $\beta$, and Wilson-Dirac operator $Q$ with hopping parameter $\kappa$, the effective theories are derived by integrating out the Grassmann fields and spatial link variables \cite{Neuman2015},
\begin{align}
	Z &= \int[\dif{U_\mu}]e^{-S_G[U_\mu]}\det Q[U_\mu] = \int[\dif{U_0}]e^{-S^\text{eff}[U_0]}\;,\notag\\
	S^\text{eff}[U_0]	& = -\text{ln} \int[\dif{U_i}] e^{-S_G[U_0,U_i]}\det Q[U_0,U_i] = S^\text{eff}_\text{G} + S_\text{stat} + S_\text{kin}^\text{eff}\;,
\end{align}
where $\det Q(U_\mu)$ is the quark determinant and $S^\text{eff}_\text{G}, S_\text{stat} $ and $ S_\text{kin}^\text{eff}$ are the gauge, static quark and kinetic quark effective action, respectively. Due to gauge invariance the effective action is a functional of Polyakov loops $S^\text{eff}[U_0] = S^\text{eff}[W]$ with $W_{\mathbf{x}}:=\prod_{\tau =0}^{N_\tau-1} U_0(\mathbf{x}, \tau)$ \cite{Langelage2011}. Because, in practice, the spatial link integration cannot be performed exactly, a combined strong coupling and hopping parameter expansion is performed as discussed in the two following subsections.
\subsection{The pure gauge effective action}
The strong coupling expansion is realised by combining a character expansion of the plaquette action with the linked cluster theorem. Then, the spatial link integrals can be performed. After deriving corrections to the leading order contribution of the expansion and applying resummation steps, the pure gauge effective action is given by \cite{Langelage2011,Fromm2012}
\begin{align}
	S_\text{G, eff} = -\sum_{\langle \mathbf{x},\mathbf{y}\rangle} \text{ln}\left(1+\lambda_1 L_{\mathbf{x}}L_{\mathbf{y}}^*+\lambda_1 L_{\mathbf{y}}L_{\mathbf{x}}^*\right)\quad \text{with}\quad  L_{\mathbf{x}} := \tr W_{\mathbf{x}}.\label{eq:pureGaugeAction}
\end{align}
The coupling $\lambda_1(\beta,N_\tau)$ depends on the couplings of the mother theory, which can be expressed analytically using the fundamental coefficients of the character expansion $u(\beta)$,~$\lambda_1(\beta,N_\tau)=\lambda_1(u(\beta),N_\tau)$. Note, that further corrections to \eqref{eq:pureGaugeAction} have been derived in previous works, implying long-range interactions and interactions between higher representations to appear in $S_\text{G}^\text{eff}$~\cite{Langelage2011}. 
\subsection{The quark effective action}
For the derivation of the quark effective action, the quark determinant $\det Q$ is separated into the static $\det Q_\text{stat}$ and the kinetic $\det Q_\text{kin}$ contribution, $\det Q  = \det Q_\text{stat} \det Q_\text{kin} $. The former can be evaluated exactly for $N_f$ degenerate quark flavours and does not depend on spatial link variables~\cite{Neuman2015},
\begin{align}
	\det Q_\text{stat} = \prod_{\mathbf{x}} \det\left(1+h_1 W_{\mathbf{x}}\vphantom{W_{\mathbf{x}}^\dagger}\right)^{2N_f}\det\left(1+\bar h_1 W_{\mathbf{x}}^\dagger\right)^{2N_f} := \prod_{\mathbf{x}}\det Q_\text{stat}^\text{loc}(W_{\mathbf{x}}) = e^{-S_\text{stat}}\;,\label{eq:statDeterminant}
\end{align}
with $h_1 = (2\kappa)^{N_\tau} e^{N_\tau a \mu}$ and $\bar h_1 = (2\kappa)^{N_\tau} e^{-N_\tau a \mu}$. 

The kinetic quark determinant describes the spatial propagation of quarks and, consequently, has to be expanded in $\kappa$ to perform the spatial link integration. This is employed by combining the trace-log identity $\det(\cdot) = \exp \tr\text{ln}( \cdot)$ with a Taylor series expansion of the exponential function and the logarithm. Afterwards, resummation techniques are applied and the kinetic quark effective action $S^{\text{eff}}_{\text{kin}}$ is obtained to the desired order in the hopping parameter. It is expressed in terms of rational functions of temporal Wilson lines,
\begin{align}
	W_{nm\bar n \bar m}(W_{\mathbf{x}}) = \tr \frac{(h_1 W_{\mathbf{x}})^m}{(1+h_1 W_{\mathbf{x}})^n}\frac{(\bar h_1 W^\dagger_{\mathbf{x}})^{\bar m}}{(1+\bar h_1 W^\dagger_{\mathbf{x}})^{\bar n}}\;, \quad W^\pm_{nm\bar n \bar m} = W_{nm00}\pm W_{00\bar n \bar m}\label{fractionalWilsonLoop}\;,
\end{align}
which can also be represented in terms of the Polyakov loop \cite{Neuman2015, Glesaaen2016}. To leading order in $\kappa$ the effective kinetic quark action $S_\text{kin, eff}$ reads \cite{Fromm2012, Neuman2015, Glesaaen2016}
\begin{align}
	S_\text{kin}^\text{eff} = 2 h_2\sum_{\langle \mathbf{x},\mathbf{y}\rangle} W_{1111}^-(W_{\mathbf{x}})W_{1111}^-(W_{\mathbf{y}})\quad \text{with}\quad h_2 = N_f N_\tau \kappa^2 /N_c\;.
\end{align}
At $\mathcal{O}(\kappa^{2n})$ the kinetic quark determinant implies couplings in the effective theories between lattice sites, that can be connected by closed paths of a Manhattan distance up to $2n$. Generally, the effective actions can be parameterised by
\begin{align}
	S^{\text{eff}}_{\text{kin}} = \sum_{\mathbf{x}} \sum_s h_s \prod_{\Delta\mathbf{x} \in s} W_{s_{\Delta \mathbf{x}}} ( W_{\mathbf{x} + \Delta\mathbf{x}}),\label{eq:kinStart}
\end{align}
where the sum over $s$ corresponds to a sum over all such closed paths and the $h_s$ are the couplings of the effective theories. The product over the $\Delta\mathbf{x}\in s$ runs over all positions on the path $s$ relative to its base point $\mathbf{x}$ and $ W_{s_{\Delta \mathbf{x}}}$ represents sums and products of the fractional Wilson lines \eqref{fractionalWilsonLoop}  \cite{Neuman2015, Glesaaen2016}.

By combining the strong coupling and hopping parameter expansion the couplings of the effective theories receive corrections and additional interaction terms appear, which are, however, always of the same form as described in this section. The effective actions are known to $\mathcal{O}(\kappa^4)$ \cite{Neuman2015} and in the cold and dense limit to $\mathcal{O}(\kappa^8)$ \cite{Glesaaen2016}.  
\section{Mean field treatment of the effective theories}
A standard technique to perform mean field approximations is realised by introducing mean fields $l$ and $\bar{l}$ for the field variables $L_{\mathbf{x}}$ and $L_{\mathbf{x}}^*$, respectively, and expanding the action around vanishing fluctuations $\delta L_{\mathbf{x}} = L_{\mathbf{x}}- l$ and $\delta L_{\mathbf{x}}^* = L_{\mathbf{x}}^*- \bar{l}$ to leading order. Alternatively, mean field approximations can be thought of as leading order saddle point approximations. There, the equations determining the saddle points are equivalent to the self-consistency equations obtained from the former technique \cite{Zinn-Justin202}. A third way of applying mean field approximations is via the Bogoliubov inequality (see e.g. \cite{Wipf2013}). As this approach relies on the probabilistic interpretation of the Boltzmann weight, the sign problem prohibits its application at non-zero chemical potential.

In the following two subsections mean field approximations will be applied to the effective theories including static and dynamical quarks, respectively.
\subsection{Mean field approximation in the static quark limit}
To present the general procedure of our mean field treatment, we consider the effective theory to leading order in the expansion parameters $\beta$ and $\kappa$, i.e. the pure gauge effective action \eqref{eq:pureGaugeAction} with the static quark determinant \eqref{eq:statDeterminant},
\begin{align}
	Z = \int [\dif{W_{\mathbf{x}}}] \prod_{\mathbf{x}}\det Q_\text{stat}^\text{loc}(W_{\mathbf{x}}) \exp\left(\sum_{\langle \mathbf{x},\mathbf{y}\rangle} \text{ln}\left(1+\lambda_1 L_{\mathbf{x}} L^*_{\mathbf{y}} + \lambda_1 L_{\mathbf{y}} L^*_{\mathbf{x}}\right)\right).\label{eq:pgStatic}
\end{align}
To perform the mean field approximation using the saddle point approximation, while treating local fluctuations exactly, the nearest-neighbour interaction is written as a power series in the $\ReN (L)$ and $\ImN(L)$ \cite{Zinn-Justin202},
\begin{align}
	\text{ln}\left(1+\lambda_1 L_{\mathbf{x}} L^*_{\mathbf{y}} + \lambda_1 L_{\mathbf{y}} L^*_{\mathbf{x}}\right) =: \sum_{n_1,m_1,n_2,m_2}\ReN(L_{\mathbf{x}})^{n_1}\ImN(L_{\mathbf{x}})^{m_1} I_{n_1,m_1,n_2,m_2} \ReN(L_{\mathbf{y}})^{n_2}\ImN(L_{\mathbf{y}})^{m_2}.
\end{align}
By introducing the auxiliary variables $\phi^{nm}_{\mathbf{x}}$ and $\Phi^{nm}_{\mathbf{x}}$ and using the relations
\begin{align}
	&\int_{-\infty}^{\infty} \dif{\phi^{nm}_{\mathbf{x}}}\delta\left(\phi^{nm}_{\mathbf{x}}-\ReN{(L_{\mathbf{x}})}^n\ImN(L_{\mathbf{x}})^m\right) \notag\\
	&= \frac{1}{2\pi i}\int_{-i\infty}^{i\infty}\dif{\Phi^{nm}_{\mathbf{x}}}\int_{-\infty}^{\infty} \dif{\phi^{nm}_{\mathbf{x}}}\exp\left({\Phi^{nm}_{\mathbf{x}}(\phi^{nm}_{\mathbf{x}}-\ReN{(L_{\mathbf{x}})}^n\ImN(L_{\mathbf{x}})^m)}\right)
\end{align}
the partition function can be rewritten exactly \cite{Zinn-Justin202},
\begin{align}
	Z &= \frac{1}{(2\pi i)^{2V}}\int[\dif{{\Phi^{nm}_{\mathbf{x}}}}]\int[\dif{\phi^{nm}_{\mathbf{x}}}]\exp\left(\sum_{\langle \mathbf{x},\mathbf{y}\rangle}\sum_{n_1,m_1,n_2,m_2}\phi^{n_1 m_1}_{\mathbf{x}} I_{n_1,m_1,n_2,m_2}\phi^{n_2 m_2}_{\mathbf{y}}\right.\notag\\
	&\left.+\sum_{\mathbf{x},n,m}\Phi^{nm}_{\mathbf{x}}\phi^{nm}_{\mathbf{x}} + \sum_{\mathbf{x}}\text{ln} z_{\mathbf{x}}\right),\\
	\text{ }&\text{with } z_{\mathbf{x}} = \int\dif{U} \det Q_\text{stat}^\text{loc}(U)\exp\left(-\sum_{n,m}\Phi^{nm}_{\mathbf{x}}\ReN{(L)}^n\ImN(L)^m\right).
\end{align}
The path integral is now performed using the saddle point approximation, where terms quadratic in non-local fluctuations around the saddle point, that maximises the partition function, are neglected. Consequently, the partition function factorises \cite{Zinn-Justin202}. The set of the infinitely many coupled saddle point equations can be summarised in a single self-consistency relation for a mean field interaction $I(\phi, \phi')$ for arbitrary values of $\phi$ and $\phi'$,
\begin{align}
	Z &= z_\text{mf}^V,\quad \text{with}\quad z_\text{mf} = e^{-d\langle I(L,L^*)\rangle_\text{mf}}\int\dif{W}\det Q_\text{stat}^\text{loc}(W) e^{2d I(L,L^*)},\\
	I(\phi,\phi') &= \left\langle \text{ln}\left(1+\lambda_1 \phi' L + \lambda_1 \phi L^*\right)\right\rangle_\text{mf},\label{eq:fSelfConsistency}
\end{align}
where $V$ is the spatial volume of the lattice and $d$ is the number of spatial dimensions. 

Similar to before one can now introduce auxiliary variables for $L$ and $L^*$, respectively, and perform another saddle point approximation, where fluctuations of the static determinant and the mean field interaction $I$ can be resummed. This allows to solve for $I(\phi, \phi')$,
\begin{align}
	I(\phi, \phi') &= \text{ln}\left(1+\lambda_1 \phi' l + \lambda_1 \phi \bar{l}\right),\label{eq:sEffPG}
\end{align}
in terms of two mean fields $l$ and $\bar{l}$, that satisfy the saddle point relations
\begin{align}
	\frac{\lambda_1 \bar{l}}{1+2\lambda_1 \bar{l} l} = \left\langle \frac{\lambda_1 L^*}{1+\lambda_1 L^* l + \lambda_1 L \bar{l}}\right\rangle_\text{mf} \text{ and }\frac{\lambda_1 l}{1+2\lambda_1 \bar{l} l} = \left\langle \frac{\lambda_1 L}{1+\lambda_1 L^* l + \lambda_1 L \bar{l}}\right\rangle_\text{mf}.\label{eq:selfconsistency}
\end{align}
The free energy density $f_\text{mf}$ is then given by
\begin{align}
	a^4 N_\tau f_\text{mf} = d\text{ln}(1+2\lambda_1 l\bar{l})-\text{ln}\int\dif{W}\det Q_\text{stat}^\text{loc}(W) \left(1+\lambda_1 L^* l + \lambda_1 L \bar{l}\right)^{2d}.\label{eq:leadingFmf}
\end{align}
Equivalently to solving the equations \eqref{eq:selfconsistency} one can now search for the saddle point minimizing the free energy density as a function of the mean fields. Further, at vanishing baryon chemical potential the saddle point relations \eqref{eq:selfconsistency} are symmetric under the exchange of the mean fields and we have $l = \bar{l}$. Instead of searching for the saddle points of the free energy one then can look for the local minima of $f_\text{mf}$ as a function of a single mean field \cite{Fukushima:2006uv, Zinn-Justin202}.

Note that the relations \eqref{eq:selfconsistency} imply, that the self-consistent mean fields are (in general) not equivalent to the expectation values of the Polyakov loop, i.e. $l \not = \langle L \rangle_\text{mf}$ and $\bar{l} \not = \langle L^* \rangle_\text{mf}$. Only when local fluctuations around the mean fields beyond leading order are neglected one recovers the usual self-consistency relations, i.e. $l  = \langle L \rangle_\text{mf} + \mathcal{O}(\delta L^2)$ and $\bar{l}  = \langle L^* \rangle_\text{mf}+ \mathcal{O}(\delta L^2)$.

Without the direct use of the saddle point approximation the equations \eqref{eq:selfconsistency} and \eqref{eq:leadingFmf} can also be derived by expressing the Polyakov loops in \eqref{eq:pgStatic} through the mean fields and the fluctuations, $L_{\mathbf{x}} = l + \delta L_{\mathbf{x}}$ and $L_{\mathbf{x}}^* = \bar{l} + \delta L_{\mathbf{x}}^*$ \cite{Wipf2013}. After neglecting terms quadratic in the fluctuations but keeping all terms that involve only local fluctuations, i.e. terms that are proportional to $\delta L_{\mathbf{x}}^n \delta L_{\mathbf{x}}^{*m}$, one obtains the desired result.
\subsection{Mean field approximation for kinetic quarks}
A natural extension to the previous subsection is the mean field approximation of the effective theories including contributions of dynamical quarks. For brevity, this approximation will here be applied with the approach described in the last paragraph of the previous subsection. For this, we start by expressing our general Ansatz for the kinetic effective quark action \eqref{eq:kinStart}  in terms of mean fields and the fluctuations around them,
\begin{align}
	S^{\text{eff}}_{\text{kin}} = \sum_{\mathbf{x}} \sum_s h_s \prod_{\Delta\mathbf{x} \in s} W_{s_{\Delta \mathbf{x}}} (l+\delta L_{\mathbf{x} + \Delta\mathbf{x}}, \bar l+\delta L_{\mathbf{x} + \Delta\mathbf{x}}^*).\label{eq:mfKinStart}
\end{align}
All the fractional temporal Wilson loops in \eqref{eq:mfKinStart} are expanded in a Taylor series around vanishing fluctuations,
\begin{align}
	W_{s_{\Delta \mathbf{x}}} (l+\delta L_{\mathbf{x} + \Delta\mathbf{x}}, \bar l+\delta L_{\mathbf{x} + \Delta\mathbf{x}}^*) &= \sum_{n,m} \frac{\delta L_{\mathbf{x} + \Delta\mathbf{x}}^n}{n!}\frac{\delta L_{\mathbf{x} + \Delta\mathbf{x}}^{*m}}{m!} \notag\\
	&\times\left.\frac{\partial^n}{\partial \delta  L_{\mathbf{x} + \Delta\mathbf{x}}^n }\frac{\partial^m}{\partial  \delta L_{\mathbf{x} + \Delta\mathbf{x}}^{*m} }W_{s_{\Delta \mathbf{x}}} (l+\delta L_{\mathbf{x} + \Delta\mathbf{x}}, \bar l+\delta L_{\mathbf{x} + \Delta\mathbf{x}}^*)\right|_{\delta L = 0}.
\end{align}
Then, terms involving fluctuations of Polyakov loops on distinct lattice sites are neglected, whereas contributions from purely local fluctuations are resummed. Consequently, the kinetic quark effective action in the mean field approximation maintains its non-linear dependence on the Polyakov loop,
\begin{align}
	S_{\text{kin}}^{\text{eff}} = \sum_{\mathbf{x}}\sum_s \left[h_s (1-|s|)\prod_{\Delta \mathbf{x} \in s} W_{s_{\Delta \mathbf{x}}}(l,\bar{l}) + \sum_{\Delta\mathbf{x}\in s}W_{s_{\mathbf{x}}}(L_{\mathbf{x}}, L_{\mathbf{x}}^*) \prod_{\substack{\Delta\mathbf{y}\in s \\\Delta \mathbf{y}\not = \Delta\mathbf{x}}}W_{s_{\Delta \mathbf{y}}}(l,\bar{l})\right] + \mathcal{O}(\delta L^2).\label{eq:mfKinFin}
\end{align}
After combining \eqref{eq:mfKinFin} with the mean field approximation of the pure gauge effective action \eqref{eq:sEffPG} and the static determinant, the self-consistent mean fields $l$ and $\bar{l}$ can be obtained by finding the saddle points of the free energy density to the desired order in the hopping parameter. 
\section{Evaluation of the effective theories using the mean field framework}
\begin{figure}[!t]
	
	\centering
	
	\includegraphics[width=1.0\linewidth]{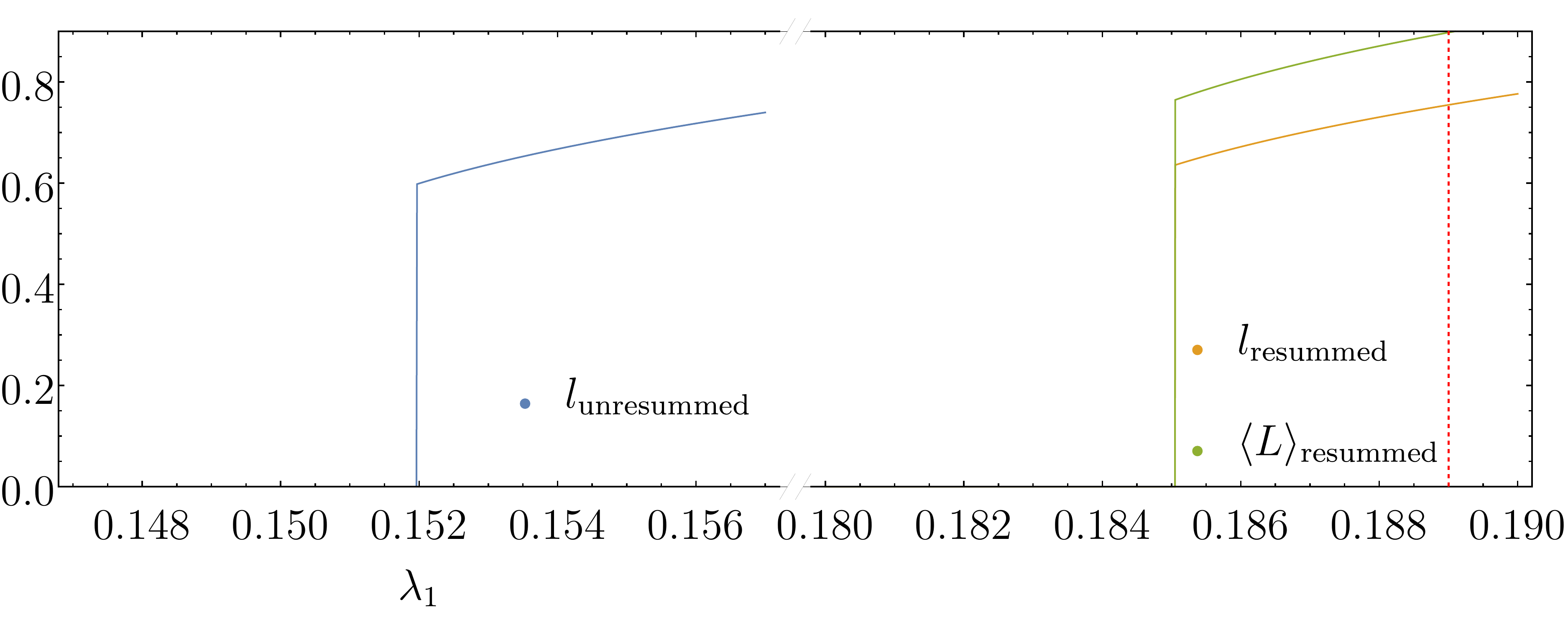}
	\captionof{figure}{The unresummed (blue line) and resummed (orange line) self-consistent mean field as well as the expectation value of the Polyakov loop (green line) versus the pure gauge effective coupling $\lambda_1$. The dashed red line signals the critical $\lambda_1$ as obtained from series expansion techniques \cite{Kim:2019ykj}.}
	\label{fig:resummedUnresummed}
	
\end{figure}
We can now test this framework in the context of the deconfinement transition. As a simple test case we consider the pure gauge limit of the effective theories and compare the effects of the mean field approximation with results, that have been obtained using series expansion techniques in reference \cite{Kim:2019ykj}.

In figure \ref{fig:resummedUnresummed} the self-consistent mean fields are shown for the resummed and non-resummed mean field approach versus the pure gauge effective coupling $\lambda_1$. There one observes, that the resummation of the fluctuations reduces the deviation of the critical coupling $\lambda_{1,c}$ from about 20\% ($\lambda_{1,c} = 0.152$) to about 3\% ($\lambda_{1,c} = 0.185$) compared to the results from series expansion techniques ($\lambda_{1,c} = 0.189$) \cite{Kim:2019ykj}. Additionally, in the same figure the expectation value of the Polyakov loop is shown for the resummed mean field approach. Due to the added corrections to the self-consistency relations the expectation value of the Polyakov loop and the self-consistent mean field are not identical any more. 
\begin{figure}[!t]
	
	\centering
	
	\includegraphics[width=0.5\linewidth]{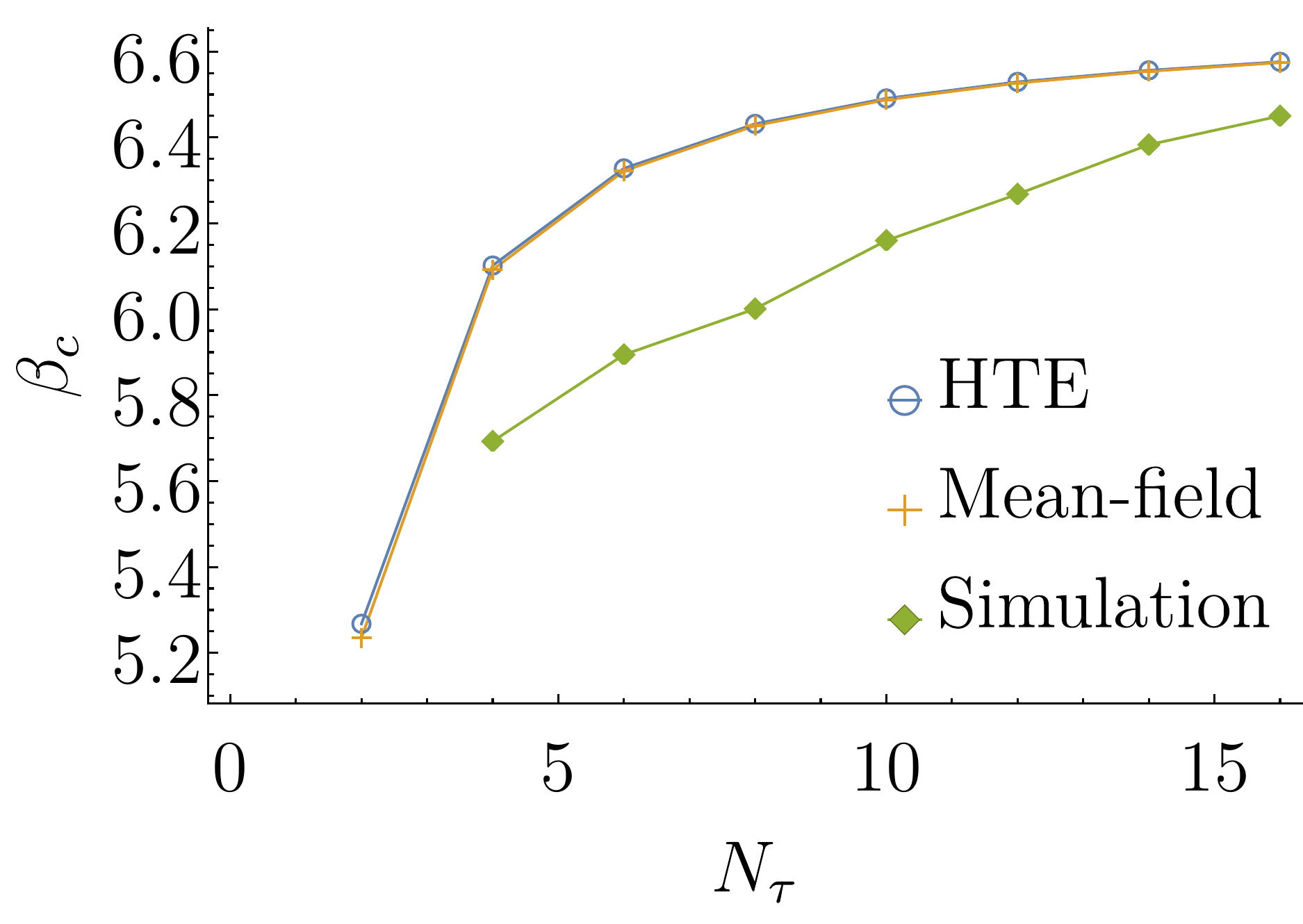}
	\captionof{figure}{The critical gauge coupling $\beta_c$ obtained from the mapping of the critical pure gauge effective coupling $\lambda_{1,c}$ for $N_\tau=2,\dots,16$ compared to results from high temperature expansion techniques (HTE)~\cite{Kim:2019ykj} and simulation results of 4D Yang-Mills theory, taken from \cite{Fingberg:1992ju}.}
	\label{fig:critBeta}
	
\end{figure}

In figure \ref{fig:critBeta} the critical gauge coupling $\beta_c$ obtained by mapping $\lambda_{1,c}$ to the coupling of the mother theory are compared to simulation results of 4D Yang-Mills theory for $N_\tau = 2,\dots,16$ \cite{Fingberg:1992ju}. There, one observes a partial cancellation of the uncertainties introduced by the derivation of the effective theory and the mean field approximation. The values of the $\beta_c$ agree with the simulation results within 10\%.
\begin{figure}[!t]\centering
	\begin{subfigure}[t]{.487\textwidth}
		\centering
		\includegraphics[width=\linewidth,page=1]{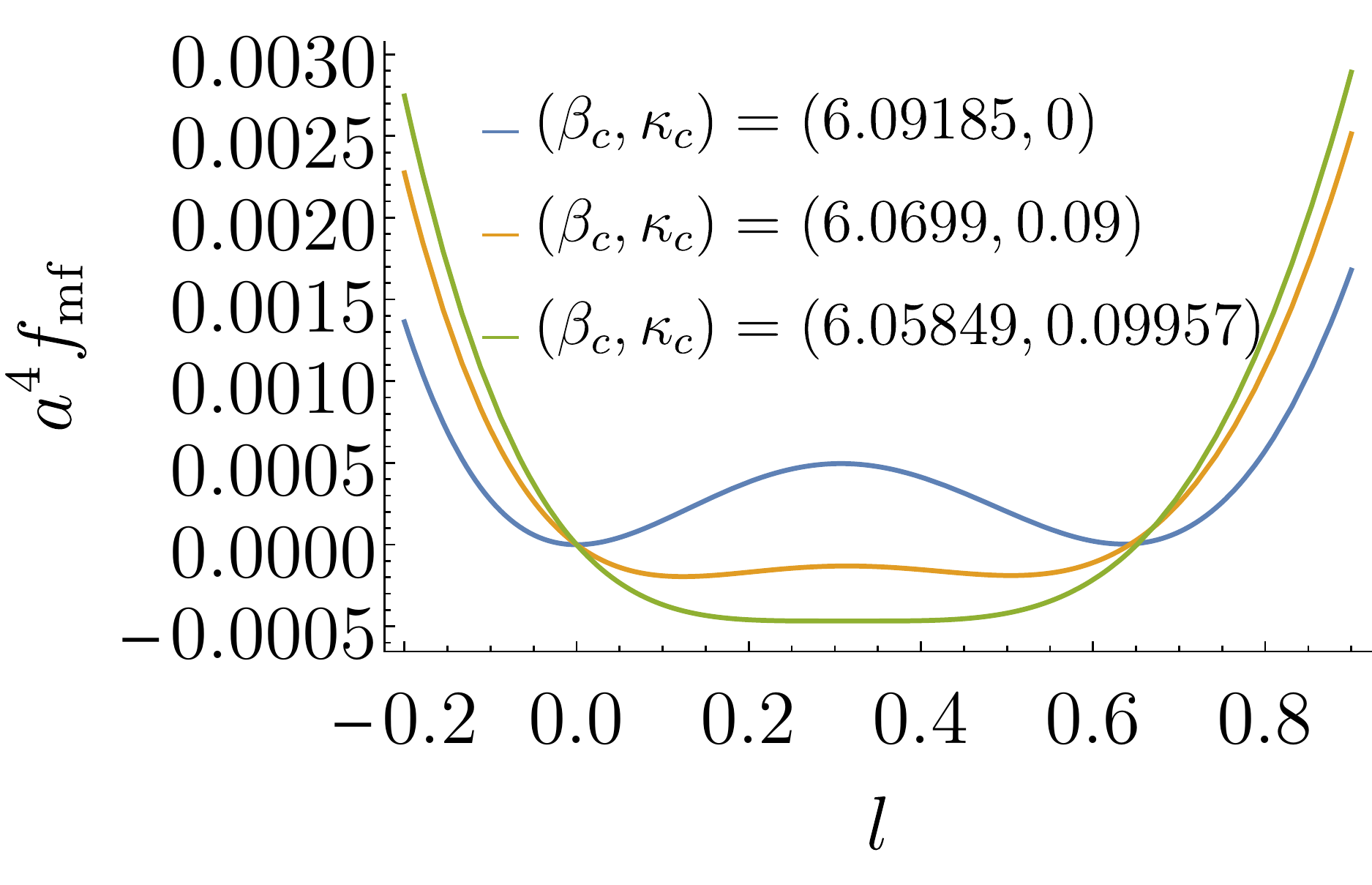}
		\caption{The free energy density $f_\text{mf}$ versus the mean field at the deconfinement transition for $N_f = 3$ and $N_\tau = 4$ in the pure gauge limit (blue line), at intermediate values of $\kappa$ (orange line) and at the critical end-point (green line).}
		\label{fig:fMf}
	\end{subfigure}%
 \quad
	\begin{subfigure}[t]{.487\textwidth}
		\centering
		\includegraphics[width=\linewidth,page=1]{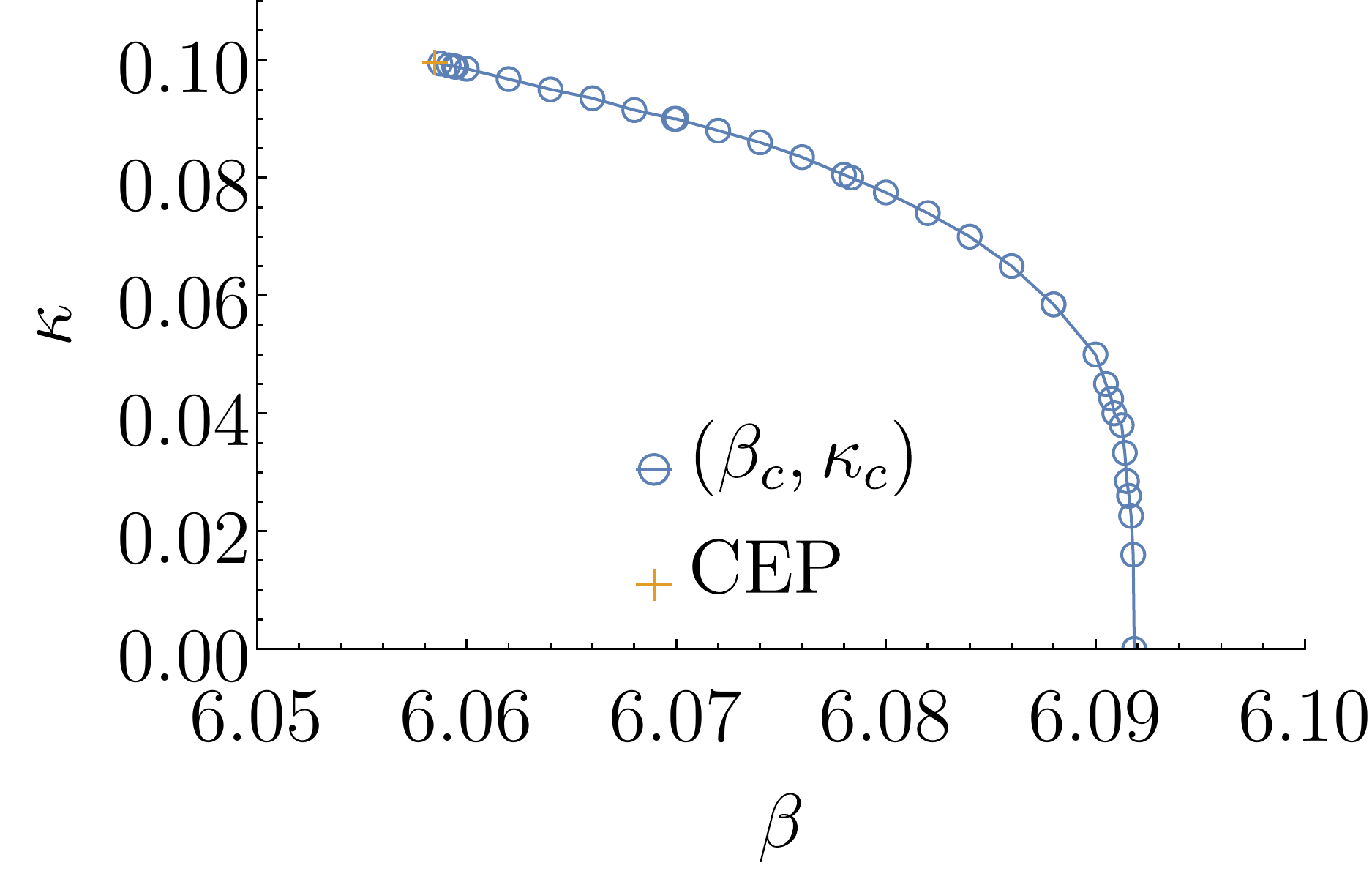}
		\caption{The deconfinement transition line (blue circles with lines as a guide to the eye) in the $\beta$-$\kappa$ plane for $N_f = 3$ and $N_\tau = 4$, starting in the pure gauge limit and ending in a second order critical end-point (CEP) (orange cross).}
		\label{fig:deconLine}
	\end{subfigure}
	\caption{The free energy density and deconfinement transition line obtained within the resummed mean field approach using the kinetic effective quark action to $\mathcal{O}(\kappa^4)$ for $N_f = 3$ and $N_\tau = 4$.}
	\label{fig:decon}
\end{figure}

Finally, we leave the pure gauge limit and consider the deconfinement transition line in the heavy quark corner for three degenerate flavours of quarks, $N_f = 3$, and $N_\tau = 4$. The kinetic quark effective action is considered to $\mathcal{O}(\kappa^4)$ as derived in \cite{Neuman2015}. In figure \ref{fig:fMf} the free energy density is shown as a function of the mean field at the deconfinement transition in the pure gauge limit (blue line), at an intermediate value of the hopping parameter (orange line), and at the critical end-point (green line), whereas in figure \ref{fig:deconLine} the deconfinement line is shown in the $\beta$-$\kappa$ plane. In both figures it can be observed, that the first order transition weakens, as the hopping parameter is increased, until it ends in a second order end-point at $\kappa_{\text{CEP}} =  0.09957$, showing a deviation of about 65\% compared to simulation results of full QCD $\kappa_{\text{CEP}} =  0.0595$ \cite{Saito:2011fs}. Such a large discrepancy for the second order end-point is expected, because mean field approximations break down when the behaviour of the system is dominated by fluctuations.
\section{Conclusions}
In this work, mean field approximations were applied to Polyakov loop effective theories of lattice QCD, that have been derived some time ago using a combined strong coupling and hopping parameter expansion. Due to the non-linear nature of the effective actions of these theories a resummation of fluctuations is used to improve the accuracy of the mean field approximation. As a showcase of the applicability of this approach the critical endpoint of the deconfinement transition is determined. The combined uncertainty of the effective theories and the mean field approximations for the critical gauge coupling ($\beta_c$) and hopping parameter ($\kappa_c$) amounts to about 15\% for first order transitions and to about 65\% for the critical end-point of the deconfinement transition. 
\section*{Acknowledgments}
The authors acknowledge support by the Deutsche Forschungsgemeinschaft (DFG, German Research Foundation) through the CRC-TR 211 'Strong-interaction matter under extreme conditions'- project number 315477589 - TRR 211 and by the State of Hesse within the Research Cluster ELEMENTS (Project ID 500/10.006).

\bibliographystyle{JHEP}
\bibliography{library}
\end{document}